\documentclass[twocolumn]{aastex62}
\usepackage{graphicx}
\usepackage[flushleft]{threeparttable}
\usepackage{blindtext}
\usepackage{ams math}
\usepackage{mathtools}
\usepackage{multirow}
\usepackage{hyperref}
\hypersetup{
	colorlinks	= true,
	linkcolor	= red,
	urlcolor	= cyan,
	citecolor	= blue
}

\newcommand{\taurex}{\mbox{$\mathcal{T}$-REx}}
\newcommand{\hatp}{\mbox{HAT-P-32\,b}}

\received{October 18, 2016}
\revised{May 10, 2017}
\accepted{May 14, 2017}
\published{July 5, 2017}

\submitjournal{AJ}

\begin{document}
	\title{Near-IR transmission spectrum of HAT-P-32\,b using HST/WFC3}
	
	\correspondingauthor{Mario Damiano}
	\email{mario.damiano.15@ucl.ac.uk}
	
	\author[0000-0002-1830-8260]{M. Damiano}
	\affiliation{Department of Physics \& Astronomy, University College London, Gower Street, WC1E6BT London, United Kingdom}
	\affiliation{INAF - Osservatorio Astronomico di Palermo, Piazza del Parlamento 1, I-90134 Palermo, Italy}
	
	\author{G. Morello}
	\affiliation{Department of Physics \& Astronomy, University College London, Gower Street, WC1E6BT London, United Kingdom}
	
	\author{A. Tsiaras}
	\affiliation{Department of Physics \& Astronomy, University College London, Gower Street, WC1E6BT London, United Kingdom}
	
	\author{T.Zingales}
	\affiliation{Department of Physics \& Astronomy, University College London, Gower Street, WC1E6BT London, United Kingdom}
	\affiliation{INAF - Osservatorio Astronomico di Palermo, Piazza del Parlamento 1, I-90134 Palermo, Italy}
	
	\author{G. Tinetti}
	\affiliation{Department of Physics \& Astronomy, University College London, Gower Street, WC1E6BT London, United Kingdom}

	\begin{abstract}
		We report here the analysis of the near-infrared transit spectrum of the hot-Jupiter \hatp\ which was recorded with the \textit{Wide Field Camera 3} (WFC3) on-board the \textit{Hubble Space Telescope} (HST). \hatp\ is one of the most inflated exoplanets discovered, making it an excellent candidate for transit spectroscopic measurements. To obtain the  transit spectrum, we have adopted  different analysis methods, both parametric and non parametric (Independent Component Analysis, ICA), and compared the results. The final spectra are all consistent within 0.5$\sigma$. The uncertainties obtained with ICA are larger than those obtained with the parametric method by a factor $\sim$1.6 - 1.8. This difference is the trade-off for higher objectivity due to the lack of any assumption about the instrument systematics compared to the parametric approach. The ICA error-bars are therefore worst-case estimates. To interpret the spectrum of \hatp\, we used \taurex, our fully Bayesian spectral retrieval code. As for other hot-Jupiters, the results are consistent with the presence of water vapor ($\log{\text{H}_2\text{O}} = -3.45_{-1.65}^{+1.83}$), clouds (top pressure between 5.16 and 1.73 bar). Spectroscopic data over a broader wavelength range will be needed to de-correlate the mixing ratio of water vapor from clouds and identify other possible molecular species in the atmosphere of \hatp.
	\end{abstract}
	

	\section{INTRODUCTION} \label{sec:introduction}
	In the past decade the \textit{Hubble Space Telescope} has been an invaluable observatory to study the properties of exoplanetary atmospheres. The majority of the planets observed to date are hot and gaseous as they are the easiest targets to probe. Transit observations in the UV, VIS and IR have started to provide important insights into the chemical composition and structure of the atmospheres of gas-giants orbiting very close to their star. Many of these atmospheres appear to be in the hydrodynamic escape regime given their vicinity to the stellar host \citep[eg.][]{Vidal2003, Linsky2010}. Common atmospheric components detected include alkali metals \citep[eg.][]{Charbonneau2002, Redfield2008} and water vapor \citep[eg.][]{Barman2007, Tinetti2007, Grillmair2008, Deming2013, Kreidberg2014b, Fraine2014}. Condensates or hazes have also been identified \citep[eg.][]{Knutson2014a, Sing2016}. Some of the data also suggest that carbon-bearing or more exotic species, such as TiO and VO \citep[eg.][]{Swain2009a, Snellen2010, Line2016, Evans2016}, are present in some of these atmospheres. Finally, eclipse and phase curve observations have enabled to glimpse into the atmospheric thermal properties and global circulation of a few of these objects \citep[eg.][]{Majeau2012, Stevenson2014}.
	
	In this work we analyze the near-infrared transit spectrum of the hot-Jupiter \hatp\ ($T_{eq} =\ 1786$\,K) \citep{2011ApJ...742...59H} obtained with the WFC3 camera on-board the HST. \hatp\ is one of the most inflated exoplanets discovered, being less massive than Jupiter ($M_\mathrm{p} = 0.79 \, M_\mathrm{Jup}$) but having almost twice its radius ($R_\mathrm{p} = 1.789 \, R_\mathrm{Jup}$). The atmosphere of \hatp\ has been observed with ground-based instruments in the optical wavelengths, revealing a featureless transmission spectrum \citep{2013MNRAS.436.2974G,2014ApJ...796..115Z, 2016arXiv160406041N, 2016arXiv160309136M}. In addition, \cite{2014ApJ...796..115Z} suggested the presence of a thermal inversion in the atmosphere of \hatp ~to interpret eclipse observations.
	
	We used our dedicated WFC3 pipeline \citep{2016ApJ...832..202T} to extract the transit light-curves per wavelength channel and obtain the planetary spectrum (Section \ref{sec:analysis}). We used in parallel Independent Component Analysis to correct for the instrumental systematics, and investigate the effect of different analysis techniques on the same data set (Section \ref{sec:ica}). The final spectrum was analyzed using our fully Bayesian spectral retrieval code, \taurex\ \citep{2015ApJ...813...13W, 2015ApJ...802..107W}.
	
	\begin{table}
		\small
		\center
		\caption{Parameters of the \hatp\ system \citep{2011ApJ...742...59H}.}
		\label{tab:parameters}
		\begin{tabular}{c | c }
			
			\hline \hline
			\multicolumn{2}{c}{Stellar parameters} \\ [0.1ex]
			\hline
			[Fe/H]\,[dex]				& -0.04 $\pm$ 0.08 \\
			$T_\mathrm{eff}$\,[K]		& 6207 $\pm$ 88	\\
			$M_* \, [M_{\odot}]$		& 1.160 $\pm$ 0.041	\\
			$R_* \, [R_{\odot}]$		& 1.219 $\pm$ 0.016	\\
			$\log(g_*)$\,[cgs] 		& 4.33 $\pm$ 0.01 \\ [1.0ex]
			
			\hline \hline
			\multicolumn{2}{c}{Planetary parameters} \\ [0.1ex]
			\hline
			$T_\mathrm{eq}$\,[K] 	& 1786$\pm$ 26 \\
			$M_\mathrm{p} \, [M_\mathrm{Jup}]$	& 0.860 $\pm$ 0.164 \\
			$R_\mathrm{p} \, [R_\mathrm{Jup}]$	& 1.789$\pm$ 0.025 \\
			$a$\,[AU] 			& 0.0343$\pm$ 0.0004 \\ [1.0ex]
			
			\hline\hline
			\multicolumn{2}{c}{Transit parameters} \\ [0.1ex]
			\hline
			$T_0$\,[BJD]			& 2454420.44637 $\pm$ 0.00009 \\
			Period\,[days] 		& 2.150008 $\pm$ 0.000001 \\
			$R_\mathrm{p}/R_*$ 	& 0.1508 $\pm$ 0.0004 \\
			$a/R_*$			& 6.05$_{-0.04}^{+0.03}$ \\
			$i$\,[deg] 			& 88.9 $\pm$ 0.4 \\ [1.0ex]
		\end{tabular}
	\end{table}
	
	\section{DATA ANALYSIS} \label{sec:analysis}
	
	\subsection{Observations} \label{sub:observation}
	
	The spatially scanned spectroscopic images of \hatp\ were obtained with the G141 grism and are available from the MAST archive\footnote{\url{https://archive.stsci.edu/}} (ID:14260, PI:Deming Drake). The data set contains five consecutive HST orbits and each exposure is the result of 14 non-destructive reads, with a size of 256$\times$256 pixels in the SPARS10 mode (exposure time = 88.435623\,s). With this configuration the maximum signal level is 2.6$\times 10^{4}$ electrons per pixel and the total scan length is approximately 40 pixels.
	
	During the light-curve analysis, the first of the five orbits was discarded. This is a standard practice for exoplanet transit observations \cite[e.g][]{2013ApJ...774...95D, 2013MNRAS.434.3252H, 2015ApJ...806..146H, 2016ApJ...820...99T}, as the telescope needs to stabilize into its new position. Of the remaining four HST orbits, the first and the fourth provide the out-of-transit baseline, while the second and the third capture the transit. The data set contains, for calibration purposes, a non-dispersed (direct) image of the target, obtained using the F139N filter.
	
	\subsection{Extraction of light-curves} \label{sub:extraction}
	
	Before extracting the light-curves (white and spectral), all frames were reduced using the routines described in \cite{2016ApJ...832..202T}. HAT-P-32\,A has an M1.5 stellar companion, HAT-P-32\,B \citep[$T_\mathrm{eff} = 3565 \pm 82 \, K$,][]{2014ApJ...796..115Z}. The dispersed signals from HAT-P-32\,A and B are blended when using the scanning mode. However, these two stars are separated enough \citep[$2^{''}.923 \pm 0^{''}.004$,][]{2014ApJ...796..115Z} to avoid blending when the differential reads (the difference between two consecutive non destructive reads, or ``stripes'') are considered. For each stripe, we determined the photometric aperture taking into account the wavelength-dependent photon trajectories \citep{2016ApJ...832..202T} and obtained a set of 12 white light-curves. The same criterion was used to extract the spectral light-curves, obtaining a set of 12 time series for each one of the 20 spectral bins. The wavelength range of each bin was chosen in order to have a similar flux level across all bins.
	
	\begin{figure}
		\centering
		\includegraphics[width=\columnwidth]{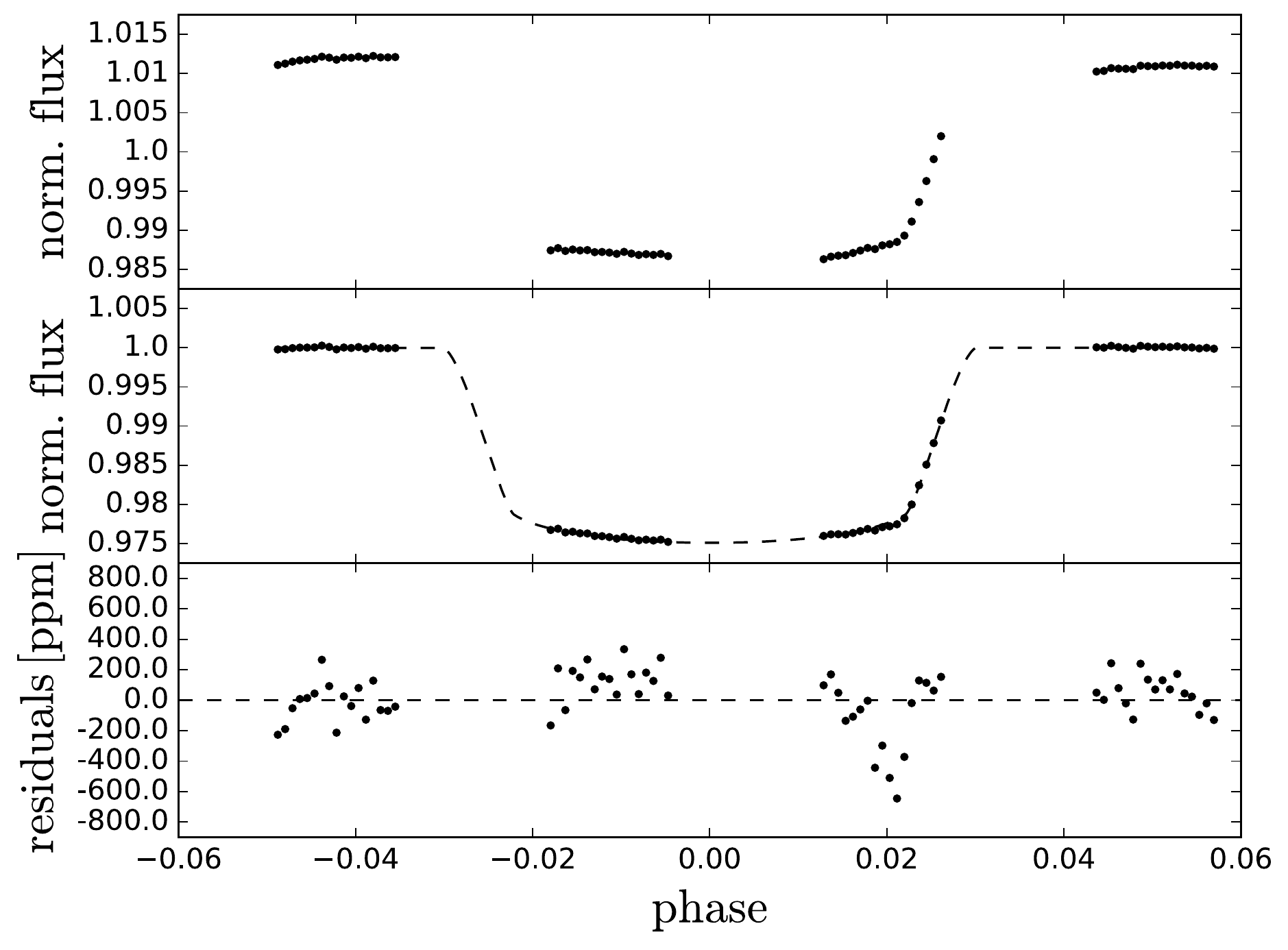}
		\caption{Top panel shows white light-curve of HAT-P-32 b. Middle panel shows fitted white light-curve. Bottom panel shows residuals after fitting.}
		\label{fig:whitelc}
	\end{figure}
	
	\subsection{Parametric fitting}\label{sub:parfitting}
	
	It is known that instrumental systematics (known as ``ramps'') affect the WFC3 infrared detector both in staring \citep{2012ApJ...747...35B, 2013Icar..225..432S, 2014ApJ...783..113W} and scanning modes \citep{2013ApJ...774...95D,2014ApJ...793L..27K, 2014Natur.505...66K, 2016ApJ...832..202T, 2016ApJ...820...99T}. The brighter is the star, the stronger are the ramps. In the case of \hatp, the host star is relatively faint ($K_\mathrm{mag} = 9.99$) so we did not expect very strong ramps. 
	
	We fitted the ramps on the white light-curve using a similar approach to \cite{2014ApJ...793L..27K}, i.e. we adopted an analytic function with two different types of ramps, short-term and long-term, to correct the data:
	
	\begin{equation}
	R(t) = (1-r_{a}(t-t_{v}))(1-r_{b1}e^{-r_{b2}(t-t_{0})})
	\label{rampfunction}
	\end{equation}
	
	\noindent where, $t$ is the mid-time of each exposure, $t_v$ is the time when the visit starts, $t_0$ is the time when each orbit starts, $r_a$ is related to the long-term ramp and $r_{b1}, r_{b2}$ are related to the short-term ramp.
	
	To model the transit light-curve we used our Python package, PyLightcurve\footnote{\url{https://github.com/ucl-exoplanets/pylightcurve}}, which returns the flux as a function of time using the non-linear limb darkening law \citep{2000A&A...363.1081C}. The limb darkening coefficients were fitted on the profile of a star similar to HAT-P-32\,A (T$_* = 6207$\,K, [Fe/H] = -0.04\,dex, $\log(g_*)$ = 4.33\,[cgs]), using a modified version of the ATLAS stellar model described in \cite{2011MNRAS.413.1515H}. In this fit we took into account the variable sensitivity of the G141 grism across its wavelength range. The observations do not cover both the ingress and the egress of the transit, hence we could not fit for the semi-major axis and inclination, which have been fixed to the values reported in Table \ref{tab:parameters}. We also assumed a circular orbit. The results are shown in Figure \ref{fig:whitelc} and reported in Table \ref{tab:fitting}. As we can see in the residuals near the egress, there are a few points that appear to deviate significantly from the model fitted to the white light-curve. The root mean square (rms) of the residuals is 180 ppm, significantly higher than the photon-noise limited rms of 100 ppm. The error bar in transit depth is accordingly higher than the photon noise limited case. This behavior could be caused by star-spots, which notoriously might generate a wavelength-dependent astrophysical signal and therefore a distortion on the spectrum. However, we repeated the analysis excluding these points and the spectrum was not affected. We also tested changing the orbital parameters up to 1$\sigma$, in both directions, from their reference values. However, this did not improve the white light-curve residuals. The main effect was a shift of $\sim$120 ppm in the while light-curve transit depth -- i.e. up to 1.3$\sigma$ from our reported uncertainty -- but again no detectable effect on the spectrum (differential transit depths vary by less than 0.25 $\sigma$ on average).
	
	\begin{table}
		\small
		\center
		\caption{White light-curve fitting results.}
		\label{tab:fitting}
		\begin{tabular}{c | c}
			
			\hline \hline
			\multicolumn{2}{c}{Limb darkening coefficients (1.125 - 1.650 $\mu$m)}	\\ [0.1ex]
			\hline	
			$a_1$ 				& 0.603336							\\
			$a_2$				& $-$0.223032							\\
			$a_3$				& 0.281379							\\
			$a_4$				& $-$0.13988							\\ [1.0ex]
			
			\hline\hline
			\multicolumn{2}{c}{Fitting results}							\\ [0.1ex]
			\hline
			$T_0 \, \mathrm{(HJD)}$			& 2457408.95783\,$\pm$\,0.00004	\\
			$R_\mathrm{p}/R_*$				& 0.1521\,$\pm$\,0.0003	
			
		\end{tabular}
	\end{table}
	
	Finally, for each wavelength bin we divided the spectral light-curve by the white light-curve \citep{2014ApJ...793L..27K} and fitted a linear trend simultaneously with a relative transit model:
	
	\begin{equation}
	n_{\lambda}(1+\chi_{\lambda})(F_{\lambda}/F_{W})
	\label{eq:methodslw}
	\end{equation}
	
	where $n_{\lambda}$ is the normalization factor that needs to be calculated for each bin, $\chi_{\lambda}$ is the wavelength-dependent linear ramp \citep{2016ApJ...832..202T, 2016ApJ...820...99T}, $(F_{\lambda}/F_{W})$ is the ratio between the spectral light-curve and the white light-curve. We fitted this model using the same orbital parameters listed in Table \ref{tab:parameters} and the white $R_{P}/R_{*}$ ratio obtained from the white light-curve fitting. The limb darkening coefficients were calculated for each bin using the same method as for the white light-curve (see Table \ref{tab:spectrum}). The rms of the residuals for the spectral light-curves (on average 474 ppm) is close to the photon noise limited case (on average 443 ppm). The corresponding error bars in relative transit depths are also $\sim$10\% above the photon noise limit. This is proves that the deviation from the model seen for the white light-curve is not wavelength-dependent.
	
	The planetary spectrum was fitted using MCMC and following two different approaches, leading to a 'stacked' and a `weighted' spectrum:
	\begin{enumerate}
		\item (stacked) using a unique reference light-curve for each spectral bin, obtained by summing the relative stripe light-curves;
		\item (weighted) fitting each stripe light-curve alone, then taking the weighted mean for each spectral bin. 
	\end{enumerate}
	Following the first method, we obtained the white light-curve shown in Figure \ref{fig:whitelc} (top panel). Both methods give the same modulation with the exception of a few bins where the differences are within 0.3 $\sigma$.
	
	\section{ICA} \label{sec:ica}
	Independent Component Analysis (ICA) is a blind signal-source separation (BSS) technique which is able to separate the source signals in a set of observations without any prior knowledge about the signals themselves or their mixing ratios. In many applications, observations are well-represented as linear combinations of certain (unknown) source signals:
	\begin{equation}\label{eqn:ICA}
	\textbf{x} = \textbf{A} \textbf{s}
	\end{equation}
	where $\textbf{x} = (x_1,x_2,...,x_n)^T$ is the column vector of observed signals, $\textbf{s} = (s_1,s_2,...,s_n)^T$ is the column vector of source signals, and $\textbf{A}$ is the so-called mixing matrix. The original source signals are retrieved through a linear transformation that maximizes their mutual independence, according to one or more statistical estimators \citep{hyvarinen00,hyvarinen12}:
	\begin{equation}\label{eqn:solveICA}
	\textbf{s} = \textbf{W} \textbf{x}
	\end{equation}
	ICA has been used to remove instrument systematics and other astrophysical signals in exoplanetary light-curves obtained with Kepler, \textit{HST}/NICMOS \citep{waldmann12,waldmann13}, \textit{Spitzer}/IRS \citep{waldmann14}, and \textit{Spitzer}/IRAC \citep{morello14,morello15,morello16,morello15b} with excellent results. We refer the reader to those publications and the relevant cited literature for more technical details about ICA and the different implementations. In this paper, we discuss a similar approach to the analysis of spectroscopic time series obtained with \textit{HST}/WFC3 using the scanning-mode technique.
	The main steps of the algorithm are:
	\begin{enumerate}
		\item ICA decomposition;
		\item Fitting;
		\item Finalizing the parameter error bars.
	\end{enumerate}
	\subsection{ICA decomposition}
	After the preliminary reduction, we obtained 12 stripe light-curves for each of the 20 spectral bins, as described in Section~\ref{sub:extraction}. We performed ICA for all bins separately, by using the corresponding 12 stripe light-curves as input time series (vector $\textbf{x}$ in Equation \ref{eqn:ICA}). Similarly, the light-curves integrated over the 20 spectral bins for each stripe were used as input white light-curves. Thus, we obtained one set of components for each spectral bin and an additional set for the whole spectral range. The transit signal is mainly contained in the first components of all sets, while the other components are predominantly instrument systematics and noise.
	\subsection{Fitting}
	Following a standard ICA algorithm (e.g. \cite{morello15b, morello16}), we simultaneously fitted a transit model (with the same parameters as in Section~\ref{sub:parfitting}) and a linear combination of the non-transit components to the relevant raw light-curves. We computed the `stacked' and a `weighted' spectra, as described in Section~\ref{sub:parfitting}.
	
	The residuals obtained for the stacked white light-curve have been included as an additional component in the spectral fits. This step is equivalent to dividing by the white light-curve as is done in the parametric fitting (see Section~\ref{sub:parfitting}), in order to remove possible undetrended systematics common to all wavelengths.  
	
	The fitting process is as follows.
	First, we run a Nelder-Mead optimization algorithm to find the parameter values minimizing the fitting residuals, then we use them as starting values for a Markov Chain Monte Carlo (MCMC) calculation with 300,000 iterations. The likelihood's variance, $\sigma_0^2$, is initialized to the variance of the residuals, then updated at any iteration. The best fitting parameters are estimated as $\mu_{par} \pm \sigma_{par,0}$, where $\mu_{par}$ and $\sigma_{par,0}$ are the mean value and standard deviation of the relevant parameter chain, respectively.
	
	\subsection{Final error bars}
	To fully account for the potential bias associated with the detrending technique, the final error bars are re-scaled with respect to the MCMC error bars inferred from the residuals only, by adding a $\sigma_{ICA}^2$ term to the likelihood's variance:
	\begin{equation}\label{eqn:final_error}
	\sigma_{par} = \sqrt{ \frac{\sigma_{ICA}^2 + \sigma_0^2}{\sigma_0^2} } \sigma_{par,0}
	\end{equation}
	The $\sigma_{ICA}^2$ term is calculated as:
	\begin{equation}\label{eqn:sigmaICA}
	\sigma_{ICA}^2 = \sum_j o_j^2 ISR_j
	\end{equation}
	where ISR is the so-called Interference-to-Signal-Ratio matrix computed with ICA, and $o_j$ are the coefficients of the non-transit components. In plain words, the $\sigma_{ICA}$ term is the weighted sum of the errors attributed to the independent components extracted with ICA. We refer the reader to \cite{morello15, morello16} for additional details.
	
	The error bars for the weighted spectrum are calculated as the simple arithmetic means of the error bars derived from fitting the independent components to the single stripes. These are worst-case estimates, as they do not scale when combining the results from the stripes. Scaling the error bars would not be theoretically correct, as the individual fits are not independent, given that they adopt the same components, which are estimated using the information contained in all the stripes. 
	The error bars obtained with ICA are larger than the ones obtained with the parametric approach by a factor $\sim$1.6 (weighted) and $\sim$1.8 (stacked). Note that, scaling the error bars in the weighted approach would have lead to final error bars smaller than photon noise limited.
	
	\section{Atmospheric retrieval}\label{sec:retrieval}
	To interpret the spectrum of \hatp, we use \taurex\ \citep{2015ApJ...802..107W, 2015ApJ...813...13W}, a Bayesian spectral retrieval code which uses line lists provided by ExoMol \citep{2012MNRAS.425...21T, 2011MNRAS.413.1828Y, 2013MNRAS.434.1469B, 2014MNRAS.440.1649Y, 2014MNRAS.437.1828B}, HITRAN \citep{2009mss..confERI01R, 2013mss..confERE03G} and HITEMP \citep{2010EGUGA..12.5561R}.
	We assumed an atmosphere dominated by  molecular Hydrogen and Helium, with a mean molecular weight of 2.3 amu. We considered as candidate trace gases a broad range of molecules, including H$_2$O, C$_2$H$_2$, CH$_4$, CO$_2$, CO, HCN, NH$_3$, VO and TiO. The RobERt (Robotic Exoplanet Recognition, \citealt{waldmann16}) module restricts the list of detectable molecules, based on the observed spectral pattern, to H$_2$O, TiO and VO. Given the relatively narrow spectral range probed, we assumed an isothermal profile and molecular abundances constant with pressure. In addition, we set uniform priors to the fitted parameters, which were: the mixing ratios of the molecules ($10^{-12}$ - $10^{-1}$), the effective temperature of the planet (1400 - 2100\,K), the radius of the planet (1.56 - 2.10\,$R_\mathrm{Jup}$) and the cloud top pressure (10$^-3$ - 10$^6$\,Pa).
	
	In addition to our best-fit model, we fitted for a fully cloudy atmosphere (straight line), and an atmosphere that contains only H$_2$O as an active gas (no TiO and VO). We calculated the Bayes factors relative to the cloudy model, as follows:
	
	\begin{equation}
	\text{B}_m = \log{\frac{\text{E}_m}{\text{E}_c}}
	\label{eq:bayes_factor}
	\end{equation}
	
	where E$_m$ is the Bayesian evidence of the test model and  E$_c$ is the one of the cloudy model.
	
	We found B$_m = 12.2$ for the pure-water model and B$_m = 12.3$ for the model that includes water, TiO and VO. These values correspond to a 5.3 $\sigma$ detection of water, while they are inconclusive about the presence of TiO and VO \citep{2008ConPh..49...71T}.

	
	\section{Results}
\begin{figure}
		\centering
		\includegraphics[width=\columnwidth]{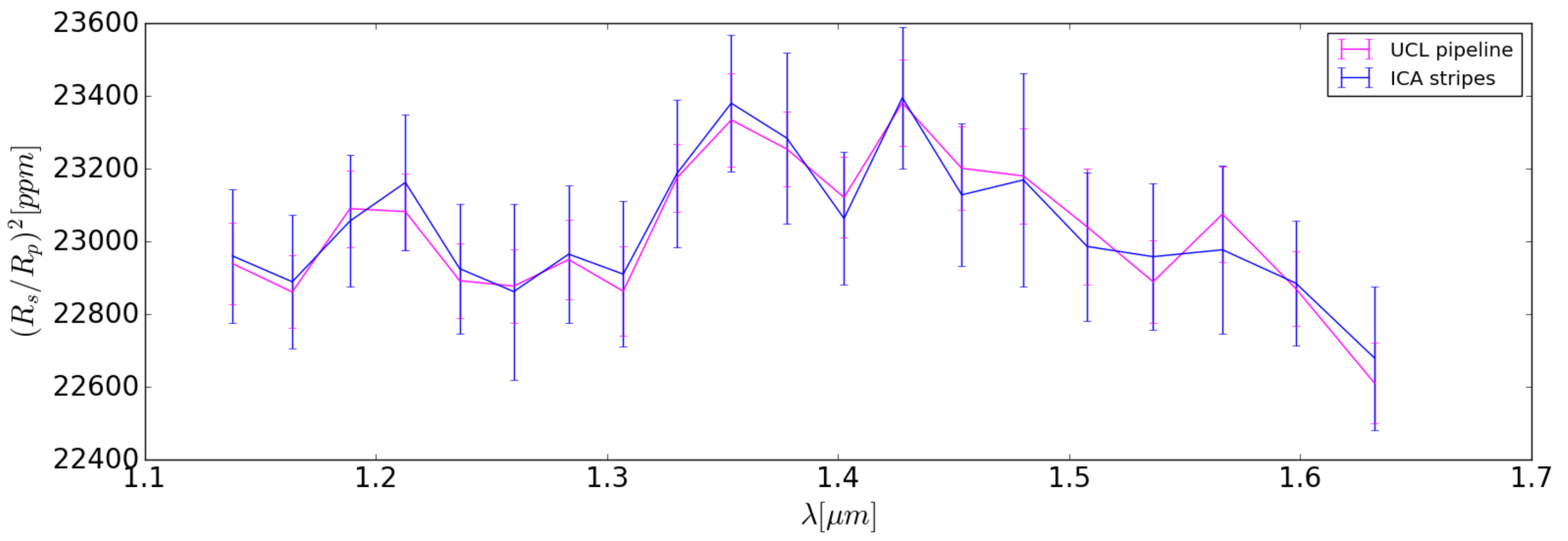}
		\caption{Stacked spectra obtained with the UCL pipeline (magenta) and with stripe-ICA (blue).}
		\label{fig:comparison}
	\end{figure}

	Figure~\ref{fig:comparison} reports the stacked spectra obtained with the parametric pipeline and with stripe-ICA. Table~\ref{tab:spectrum} reports the numerical results. The four spectra, i.e. stacked and weighted obtained with the two detrending algorithms, are all consistent within 0.5~$\sigma$ (maximum discrepancy for a wavelength-bin). Also, the error bars for the corresponding stacked and weighted spectra are similar, within less than 10$\%$ in average.
	
	\begin{table}
		\small
		\center
		\caption{Fitting results for HAT-P-32 b atmosphere}
		\label{tab:fitting_results}
		\begin{tabular}{c | c }
			
			\hline \hline
			Parameter	& Value											\\ [0.1ex]
			\hline
			$\log{\text{H}_2\text{O}}$	& $-4.66_{-1.93}^{+1.66}$				\\
			$T_\mathrm{eff} \, \mathrm{[K]}$		& $1553_{-91}^{+174}$		\\
			$R_p \, [R_{\mathrm{jup}}]$				& $-1.76_{-0.04}^{+0.05}$	\\
			$P_\mathrm{cld, top} \, \mathrm{[bar]}$	& $3.39_{-1.66}^{+1.77}$	\\ 
		\end{tabular}
	\end{table}
	
	\begin{table*}
		\small
		\center
		\caption{Limb darkening coefficients $a_{1-4}$ and transit depth $(R_p/R_*)^2$ for the wavelength channels.}
		\label{tab:spectrum}
		\begin{tabular}{r l | c c c c | c c}
			\hline \hline
			\multicolumn{2}{c |}{$\lambda_1- \lambda_2 \, (\mu \mathrm{m})$}	 & \multirow{2}{*}{$a_1$}	 & \multirow{2}{*}{$a_2$} 	 & \multirow{2}{*}{$a_3$}	 & \multirow{2}{*}{$a_4$} 	 & $(R_\mathrm{p}/R_*)^2 \, \mathrm{(ppm)}$ 	& $(R_\mathrm{p}/R_*)^2 \, \mathrm{(ppm)}$ \\ [0.1ex]
			& & & & & & UCL pipeline & stripe-ICA\\
			\hline
			1.1250 	 & 1.1511 					 & 0.632741 	 & -0.481904 	 & 0.701108 	 & -0.306091 	 & 22940 $\pm$ 112 	& 22961 $\pm$ 184\\
			1.1511 	 & 1.1767 					 & 0.619205 	 & -0.434713 	 & 0.64011 	 & -0.282483 	 & 22862 $\pm$ 100 	& 22890 $\pm$ 184\\
			1.1767 	 & 1.2011 					 & 0.614294 	 & -0.41589 	 & 0.610565 	 & -0.272242 	 & 23091 $\pm$ 105 	& 23057 $\pm$ 181\\
			1.2011 	 & 1.2247 					 & 0.599151 	 & -0.360648 	 & 0.544934 	 & -0.247917 	 & 23083 $\pm$ 105 	& 23163 $\pm$ 186\\
			1.2247 	 & 1.2480 					 & 0.584001 	 & -0.29953 	 & 0.465487 	 & -0.216442 	 & 22893 $\pm$ 102 	& 22926 $\pm$ 179\\
			1.2480 	 & 1.2716 					 & 0.581928 	 & -0.282551 	 & 0.441745 	 & -0.210655 	 & 22878 $\pm$ 102 	& 22863 $\pm$ 242\\
			1.2716 	 & 1.2955 					 & 0.58946 	 & -0.229732 	 & 0.322997 	 & -0.169253 	 & 22951 $\pm$ 110 	& 22966 $\pm$ 189\\
			1.2955 	 & 1.3188 					 & 0.57237 	 & -0.227002 	 & 0.362724 	 & -0.181489 	 & 22864 $\pm$ 123 	& 22911 $\pm$ 200\\
			1.3188 	 & 1.3421 					 & 0.569522 	 & -0.202303 	 & 0.325228 	 & -0.166816 	 & 23176 $\pm$ 94 	& 23188 $\pm$ 203\\
			1.3421 	 & 1.3657 					 & 0.564634 	 & -0.163366 	 & 0.265035 	 & -0.14235 	 & 23335 $\pm$ 129 	& 23381 $\pm$ 189\\
			1.3657 	 & 1.3901 					 & 0.561817 	 & -0.127278 	 & 0.200548 	 & -0.113503 	 & 23255 $\pm$ 103 	& 23285 $\pm$ 236\\
			1.3901 	 & 1.4152 					 & 0.561832 	 & -0.0979712 	 & 0.148201 	 & -0.0914278 	 & 23122 $\pm$ 111 	& 23064 $\pm$ 182\\
			1.4152 	 & 1.4406 					 & 0.572262 	 & -0.100901 	 & 0.133369 	 & -0.0848254 	 & 23382 $\pm$ 119 	& 23396 $\pm$ 195\\
			1.4406 	 & 1.4667 					 & 0.58462 	 & -0.111943 	 & 0.124656 	 & -0.0799948 	 & 23202 $\pm$ 115 	& 23129 $\pm$ 196\\
			1.4667 	 & 1.4939 					 & 0.600205 	 & -0.136878 	 & 0.140204 	 & -0.0874595 	 & 23181 $\pm$ 130 	& 23170 $\pm$ 294\\
			1.4939 	 & 1.5219 					 & 0.609784 	 & -0.134319 	 & 0.11158 	 & -0.0721681 	 & 23041 $\pm$ 160 	& 22987 $\pm$ 204\\
			1.5219 	 & 1.5510 					 & 0.626375 	 & -0.139701 	 & 0.0839621 	 & -0.0555132 	 & 22890 $\pm$ 114 	& 22959 $\pm$ 201\\
			1.5510 	 & 1.5819 					 & 0.647904 	 & -0.193435 	 & 0.120068 	 & -0.0635888 	 & 23076 $\pm$ 131 	& 22978 $\pm$ 230\\
			1.5819 	 & 1.6145 					 & 0.663831 	 & -0.223633 	 & 0.124246 	 & -0.0583813 	 & 22871 $\pm$ 102 	& 22886 $\pm$ 171\\
			1.6145 	 & 1.6500 					 & 0.686226 	 & -0.267069 	 & 0.137329 	 & -0.0557593 	 & 22611 $\pm$ 111 	& 22680 $\pm$ 198\\
		\end{tabular}
	\end{table*}
	
	\subsection{Retrieval Results}
\begin{figure*}
		\centering
		\includegraphics[width=0.8\textwidth]{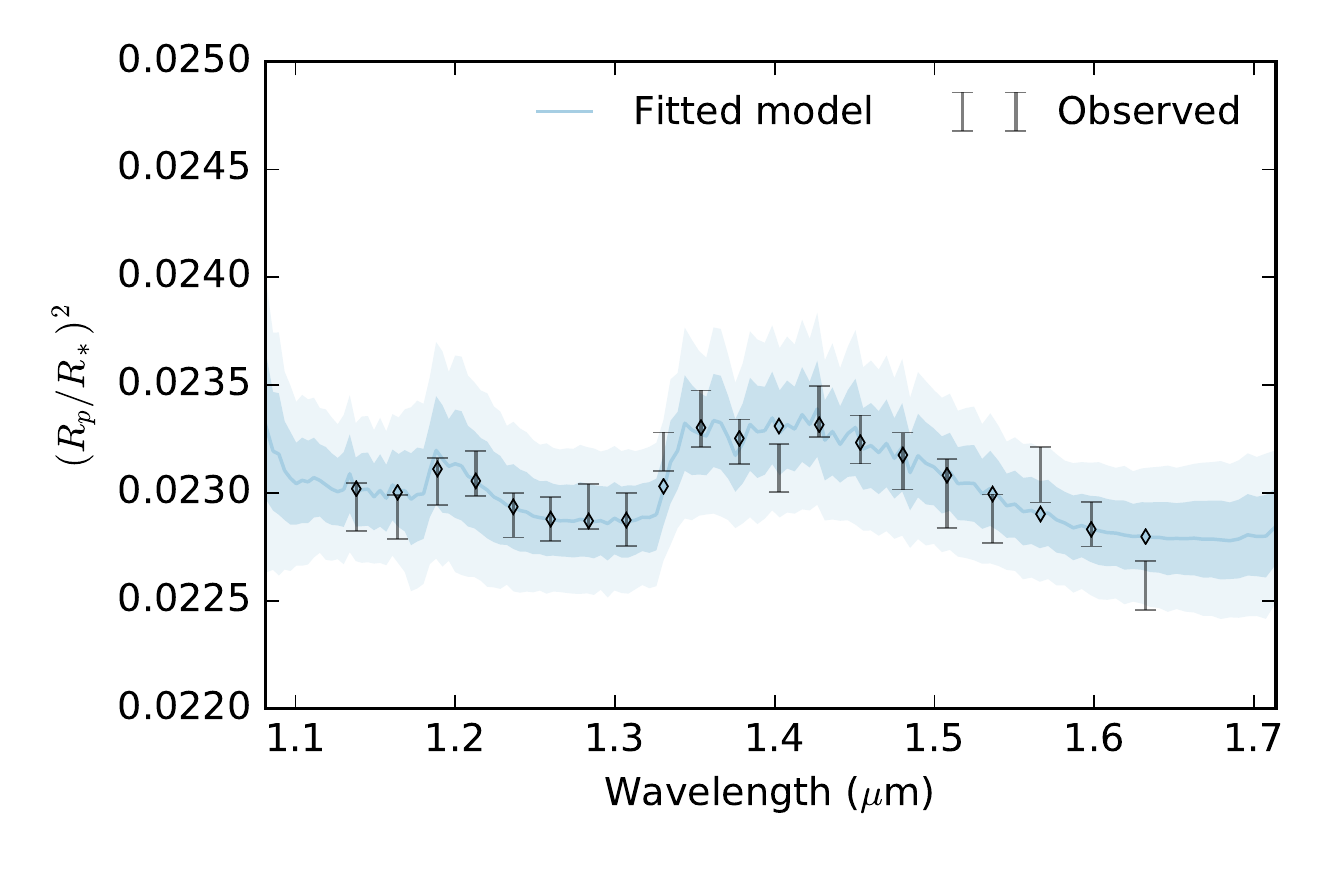}
		\caption{Transmission spectrum of HAT-P-32 b obtained with the UCL pipeline (black) and best fitting model (light-blue).}
		\label{fig:binned_spectrum}
	\end{figure*}

\begin{figure}
		\centering
		\includegraphics[width=\columnwidth]{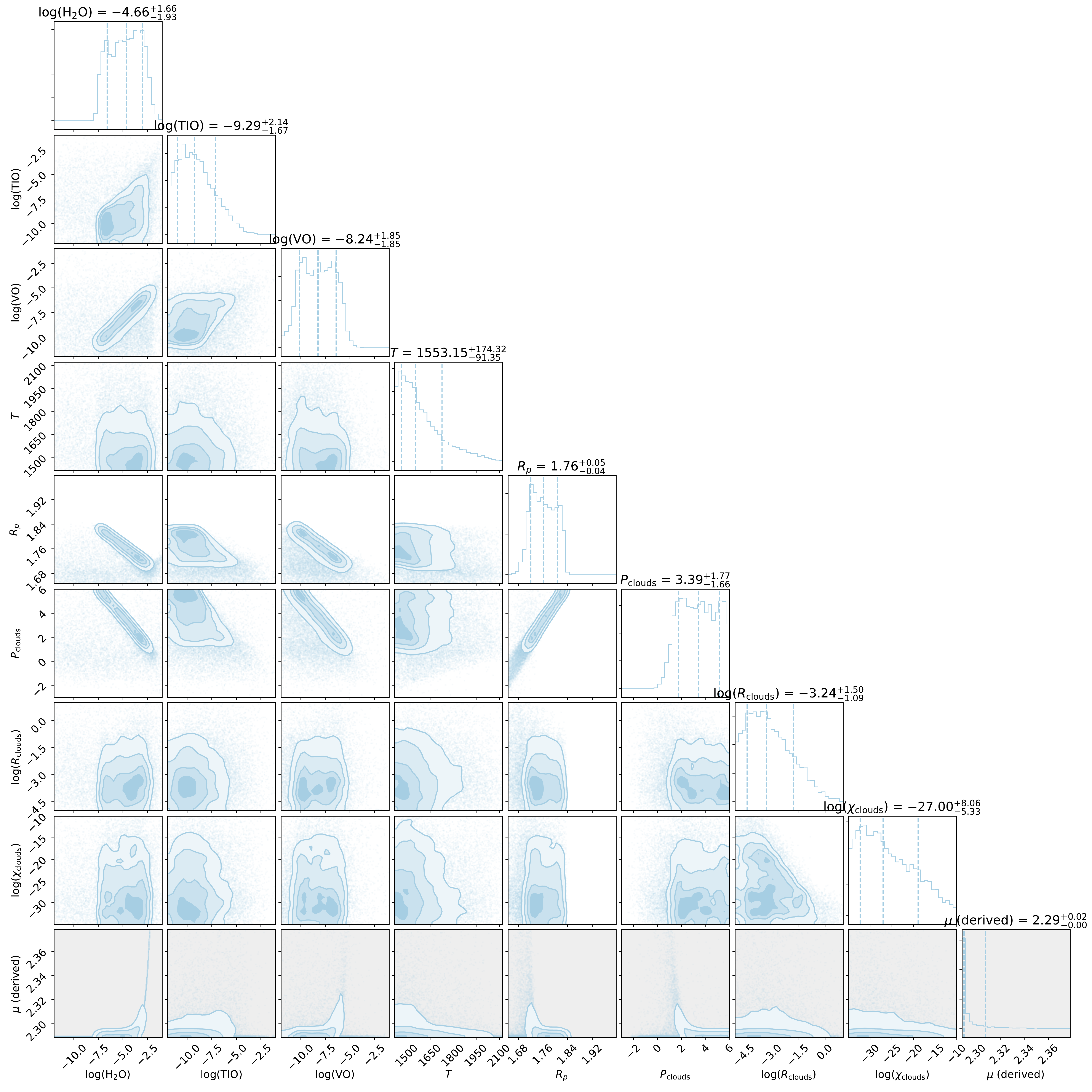}
		\caption{Posterior distributions to the fit for the WFC3 spectrum of the giant planet HAT-P-32 b. Even though we tested the presence of many other molecules in this atmosphere, here we show only the posterior of H$_2$O because it is the only significant one. All the other molecules do not show a statistically significant contribution to the fit.}
		\label{fig:posteriors}
	\end{figure}

	The transmission spectrum of HAT-P-32 b and the best fit to it, retrieved with \taurex, is shown in Figure \ref{fig:binned_spectrum}. The best fitting values and the posterior distributions are shown in Tab \ref{tab:fitting_results} and Figure \ref{fig:posteriors}.
	With the  exception of water vapor, the fitted values for all the other molecular mixing ratios are smaller than $10^{-7}$. This result means that they are not detectable from this data-set. The water vapor mixing ratio oscillates, instead, between  $\log{\text{H}_2\text{O}} = -3.45_{-1.65}^{+1.83}$ depending on the clouds' top pressure, which could occur between 5.16 and 1.73 bar. A strong correlation between water vapor mixing ratio, clouds' top pressure, planetary radius at 10 bar and temperature is noticeable in Figure \ref{fig:posteriors}, indicating there is a degeneracy of solutions.
	
	\section{DISCUSSION} \label{sec:discussion}
	\subsection{Comparison between detrending algorithms} \label{sub:comparison1}
	As mentioned in the previous section, the error-bars obtained with ICA are larger by a factor $\sim$1.6 - 1.8 compared to the ones obtained with the parametric fitting. The larger-error bars obtained with ICA are the trade-off for higher objectivity, due to the lack of any assumption about the instrument systematics compared to the parametric approach. The ICA error-bars are worst-case estimates. It is worth noting that the discrepancies between the spectra obtained with the different methods are smaller than the parametric error-bars, suggesting that, in this case, the ICA error-bars might be overly conservative.
	
	\subsection{Comparison with other observations} \label{sub:comparison2}
	
	Previous ground-based observations of the transit of \hatp\ in the optical wavelengths \citep{2013MNRAS.436.2974G,2014ApJ...796..115Z, 2016arXiv160406041N, 2016arXiv160309136M} did not find evidence of spectral modulations due to molecules. Our cloud top pressure is consistent with their measurements within 1$\sigma$, hence the water detection in the infrared is not controversial.
	
	\subsection{Strong water feature} \label{sub:waterfeature}
	
	Water vapour has been detected, to date, in the atmospheres of about 10 hot Jupiters \citep{iyer16}. \cite{2016ApJ...817L..16S} identify two classes of hot Jupiters, essentially mostly cloudy or with a strong water signature. The observed trend suggests that hotter ($T_{eq} > \ 700$\,K) and more inflated ($\log{g} >\ 2.8$) planets are more likely to have a strong water signature than cooler and smaller ones, but the current sample is not statistically significant. In agreement with this scenario, we find that \hatp\ ($T_{eq} =\ 1786$\,K; $\log{g} >\ 2.8$) has one of the strongest water features so far detected ($\sim$ 500 ppm, 5.3 $\sigma$).
	
	\section{CONCLUSION} \label{sec:conclusion}
	We have reported here the analysis of the near-infrared transit spectrum of the hot-Jupiter HAT-P-32 b which was recorded with the  \textit{Wide Field Camera 3} on-board the \textit{Hubble Space Telescope}. 
	
	To obtain the  transit spectrum, we have adopted different analysis methods, which include  parametric and non parametric techniques (Independent Component Analysis, ICA), and compared the results. The final spectra are all consistent within 0.5$\sigma$. The uncertainties obtained with ICA are larger than the ones obtained with the parametric method by a factor $\sim$1.6 - 1.8. The larger uncertainties obtained with ICA are the trade-off for higher objectivity, due to the lack of any assumption about the instrument systematics compared to the parametric approach. The ICA uncertainties are therefore the worst-case estimates. 
	
	To interpret the spectrum of HAT-P-32 b, we used \taurex, a fully Bayesian spectral retrieval code.

	As for other hot-Jupiters, the results are consistent with the presence of water vapor ($\log{\text{H}_2\text{O}} = -4.66_{-1.93}^{+1.66}$) and probably clouds (top pressure between 5.16 and 1.73 bar). Spectroscopic data over a broader wavelength range will be needed to de-correlate water vapour's mixing ratio from clouds and identify other possible molecular species in HAT-P-32 b atmosphere. 
	
	\section{Ackowledgements} 
	The authors are supported by the European Research Council (ERC) grant \emph{ExoLights}. M. D. and T. Z. are also supported by the Istituto Nazionale di AstroFisica - Osservatorio Astronomico di Palermo (INAF-OAPa). G. T. is supported by a Royal Society URF. This work is also supported by STFC (ST/P000282/1).
	
	\noindent The authors wish to thank Ingo Waldmann, Marco Rocchetto, Jonathan Tennyson and Sergey Yurchenko for their input.
	
	{\small
		\bibliography{wfc3_hat_p_32b}{}
		\bibliographystyle{apj}
	}
	
	\newpage
	
	\appendix
	
	\section{Dilution factor for the companion star}
	
	In this particular data set the nearby companion HAT-P-32B can be separated from the host star HAT-P-32A. This gives us the opportunity to calculate the dilution factor between the two stars , so that it can be used by other studies where the two stars cannot be separated. To calculate the dilution factor we extracted the light-curve of the companion by adjusting the wavelength calibration and using an aperture expanding five pixels above and below the spatially scanned spectrum. Given the undispersed image of the system taken in the beginning of the observation, we found the companion to be shifted by -3.5487 and -23.9709 pixels along the horizontal and vertical axes of the detector, respectively. For this calculation we considered only the spectra obtained during the last HST orbit as we noticed that the dilution factor was varying linearly during the two orbits before the transit. The most possible explanation for this behavior is the long-term, linear, ramp.
	
	\begin{table}[h!]
		\small
		\center
		\caption{Dilution factor for the nearby companion in each wavelength channel.}
		\label{tab:dilution}
		\begin{tabular}{c c | c}
			\hline \hline
			\multicolumn{2}{c |}{$\lambda_1- \lambda_2 \, (\mu \mathrm{m})$} & dilution factor \\ [0.1ex]
			\hline
			1.1250 & 1.1511 & 0.03232 $\pm$ 0.00066 \\
			1.1511 & 1.1767 & 0.03272 $\pm$ 0.00068 \\
			1.1767 & 1.2011 & 0.03297 $\pm$ 0.00069 \\
			1.2011 & 1.2247 & 0.02607 $\pm$ 0.00043 \\
			1.2247 & 1.2480 & 0.02756 $\pm$ 0.00048 \\
			1.2480 & 1.2716 & 0.02845 $\pm$ 0.00051 \\
			1.2716 & 1.2955 & 0.02960 $\pm$ 0.00056 \\
			1.2955 & 1.3188 & 0.03050 $\pm$ 0.00059 \\
			1.3188 & 1.3421 & 0.03312 $\pm$ 0.00069 \\
			1.3421 & 1.3657 & 0.03318 $\pm$ 0.00070 \\
			1.3657 & 1.3901 & 0.03303 $\pm$ 0.00069 \\
			1.3901 & 1.4152 & 0.03307 $\pm$ 0.00069 \\
			1.4152 & 1.4406 & 0.03167 $\pm$ 0.00063 \\
			1.4406 & 1.4667 & 0.04220 $\pm$ 0.00113 \\
			1.4667 & 1.4939 & 0.04108 $\pm$ 0.00107 \\
			1.4939 & 1.5219 & 0.03937 $\pm$ 0.00098 \\
			1.5219 & 1.5510 & 0.03738 $\pm$ 0.00088 \\
			1.5510 & 1.5819 & 0.03602 $\pm$ 0.00082 \\
			1.5819 & 1.6145 & 0.03478 $\pm$ 0.00076 \\
			1.6145 & 1.6500 & 0.03412 $\pm$ 0.00074 \\
		\end{tabular}
	\end{table}
	
	\newpage

\end{document}